# Unconventional spin polarization at Argon ion milled SrTiO$_3$ Interfaces.

Amrendra Kumar[1], Utkarsh Shashank[2], Suman Kumar Maharana[1], John Rex Mohan[2], Surbhi Gupta[3], Hironori Asada[4], Yasuhiro Fukuma[2,5]* and Rohit Medwal[1]*

[1]Department of Physics, Indian Institute of Technology Kanpur, Kanpur 208016, India

[2]Department of Physics and Information Technology, Faculty of Computer Science and Systems Engineering, Kyushu Institute of Technology, 680-4 Kawazu, Iizuka 820-8502, Japan

[3]Department of Physics, Motilal Nehru National Institute of Technology, Barrister Mullah Colony, Teliarganj, Prayagraj, Uttar Pradesh, 211004, India

[4]Department of Electronic Devices and Engineering, Graduate School of Science and Engineering, Yamaguchi University

[5]Research Center for Neuromorphic AI hardware, Kyushu Institute of Technology, Kitakyushu 808-0196, Japan

*Correspondence should be addressed to R.M.: rmedwal@iitk.ac.in

Co-correspondence should be addressed to Y.F..: fukuma@phys.kyutech.ac.jp

**ABSTRACT**

Interfacial two-dimensional electron gas (2DEG) formed at the perovskite-type oxide, such as SrTiO3, has attracted significant attention due to its properties of ferromagnetism, superconductivity, and its potential application in oxide-based low-power consumption electronics. Recent studies have investigated spin-to-charge conversion at the STO interface with different materials, which could affect the efficiency of this 2DEG interface. In this report, we presented a $Ar^+$ ion milling method to create a 2DEG at STO directly by inducing oxygen vacancies. To quantify the spin-to-charge conversion of this interface, we measured the angular-dependent spin torque ferromagnetic resonance (ST-FMR) spectra, revealing an unconventional spin polarization at the interface of of Argon ion milled STO and NiFe. Furthermore, a micromagnetic simulation for angular dependent spin torque ferromagnetic resonance (ST-FMR) has been performed, confirming the large unconventional spin type spin polarization at the interface.

# I. INTRODUCTION

Efficient charge to spin conversion at interface has emerged as a promising research direction in the field of spintronics for the development of next-generation spin based electronic devices[1,2,3]. Unlike conventional electronics, which rely solely on the charge of electrons, spintronics exploit both the charge and the intrinsic spin of electrons to manipulate and store information at high speed and low power consumption. In recent years it has been found out that there are two major mechanisms for the conversion of charge to spin in nonmagnetic materials using spin hall effect and Rsahba-Edelstein[4,5,6,7] effect which are called Bulk effect and Interface effect respectively. A crucial element in spintronics research is the two-dimensional electron gas (2DEG)[8,9,10,11], a unique physical system that plays a pivotal role in enabling numerous spintronic applications. Therefore, 2DEG is a foundational component in spintronics research, offering a versatile platform for investigating and harnessing the spin properties of electrons. Its unique properties, such as spin-polarized transport, efficient spin injection and detection, and strong spin Hall effect, make it a crucial building block for various spintronic applications, including quantum computing, spin-FETs, and spin filters. The ongoing research and development in 2DEG[12,13,14] based spintronics holds great promise for the advancement of future electronic and computing technologies.

It has been reported that perovskite oxide-based interfaces have shown formation of 2D electron gas as STO/ALO[39]. Mainly STO has its 2D electron gas at its surface due to the vacant oxygen sites. In this work, we present the formation of 2D electron gas using $Ar^+$ ion milling by removal of oxygen from STO surfaces. This formed 2D electron gas shows inversion symmetry at the interface, creating Rashba effect at the surface of STO resulting in out-of-plane electric field. So, when the charge current $J_x$ flows through the surface, it generates a transverse spin current through Edelstein effect. which upon diffusing to adjacent ferromagnetic materials exerts a torque on its magnetization.

**Experimental:**

The $SrTiO_3$ single crystal possesses a unique conducting surface state, which make it ideal for applications involving spin-dependent transport[15,16,17]. Additionally, $SrTiO_3$ shows the large spin polarization due to strong spin-orbit coupling, unique electronic band structure, and quantum confinement of the 2DEG. The $SrTiO_3$[15,16,17] single crystal can host an unusual conducting surface state which can be used for spin dependent charge transport and large spin polarization. To develop the conducting surface of $SrTiO_3$, an ion milling process was used, where an ion gun produced a beam of high energetic ions (typically noble gases like Ar+ ions) and bombarded its towards the surface of crystal. To create conducting surface of $SrTiO_3$, accelerated ions, also called ion milling processes, were used. This bombardment of Argon ions on the surface of $SrTiO3$ with orientation (100) builds a conducting surface state by removing oxygen ions from $SrTiO_3$. The choice of inert gas helps us to prevent the $SrTiO_3$, from additional impurities. It is expected that the thickness of the surface conducting layer changes with increasing the milling time. This can be simply co-related with the conductance of the surface as a function of milling time. After confirming the formation of conducting surface states, ferromagnetic layer $Ni_{80}Fe_{20}$ (hereafter refer as $NiFe$) was deposited on ion milled $SrTiO_3$. The designed $SrTiO_3/2DEG/NiFe$ stack were subjected to micro device fabrication using photolithography technique for spin transport.

To perform the spin dependent transport measurements, spin torque ferromagnetic resonance (ST-FMR)[18,19] devices were prepared in the in-plane excitation geometry as shown in Fig. 1(b). In the ST-FMR measurement, a microwave current $I_{rf}$ is applied in the longitudinal direction along with an in-plane external magnetic field $H_{ext}$ at an angle ($\phi$) of 0° to the sample. The applied $I_{rf}$ current generates Oersted field (Ampere's law) which thereby exerts an Oersted field torque ($\tau_{Oe}$) on magnetization vector of $NiFe$ layer while injected $I_{rf}$ into 2DEG layer will get convert into oscillating transverse spin current. This results in a spin orbit torque ($\tau_{Oe}$) acting on the magnetization vector of $NiFe$. Thus, both Oersted field torque and spin orbit torque collectively drives the magnetization precession in $NiFe$ along the effective magnetic field ($H_{eff}$) as shown in Fig. 1(b), when applied microwave frequency and external magnetic field satisfy resonance condition. This precession leads to oscillation of resistance due to anisotropic magnetoresistance (AMR) of $NiFe$.

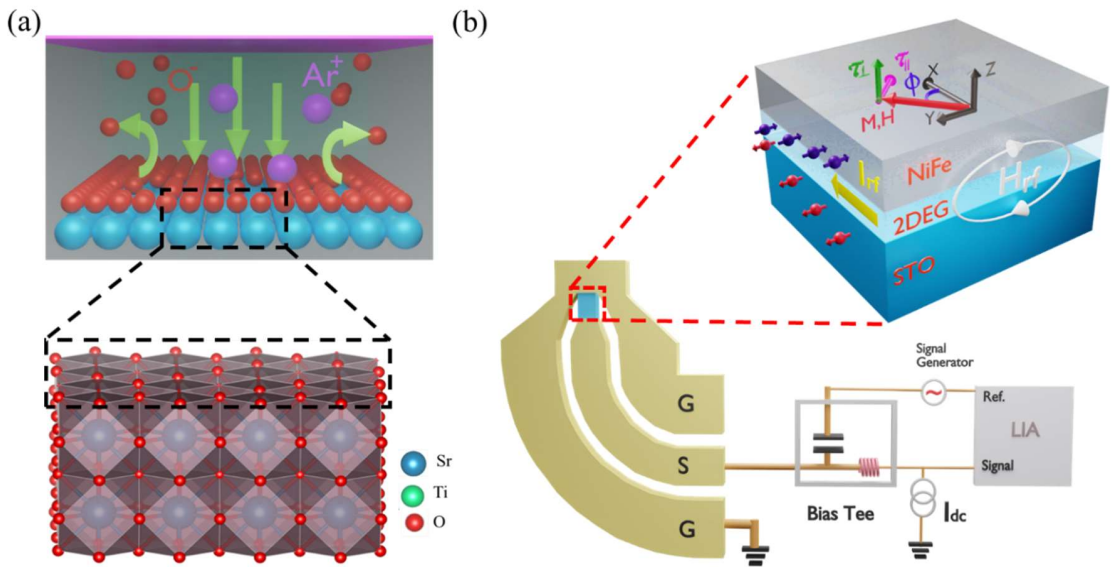

*Figure 1. (a) Schematic showing ($Ar^+$) ion milling in SrTiO₃ substrate (b) Schematic of ST-FMR measurement with lock-in setup to measure the $V_{mix}$ voltage across the device.*

A rectified DC voltage across the SrTiO₃/2DEG/NiFe device from the mixing of $I_{rf}$ and oscillating resistance is detected using lock-in and high frequency bias tee, while applying simultaneously the microwave. The rectifying voltage $V_{mix}$ was measured using phase sensitive lock in technique while sweeping an external in plane magnetic field was swept at an angle of $\phi$. The rectifying voltage $V_{mix}$ is fitted using the following equation[20,21],

$$V_{mix} = V_S F_{Sym}(H_{ext}) + V_A F_{Asym}(H_{ext}), \tag{1}$$

where, $V_{Sym}(H_{ext}) = \frac{(\Delta H)^2}{(H_{ext}-H_{res})^2+(\Delta H)^2}$, is the symmetric part of the $V_{mix}$ spectrum, $F_{Asym}(H_{ext}) = \frac{\Delta H(H_{ext}-H_{res})}{(H_{ext}-H_{res})^2+(\Delta H)^2}$, is the antisymmetric part, $\Delta H$ and $H_{res}$ are the half-width-at-half-maximum (linewidth) and the resonance field, and amplitude $V_S$ which is associated with the symmetric part of the spectra and related to in plane torques($\tau_\parallel$), and $V_A$ related to antisymmetric part of the spectra which is due to out-of-plane torques ($\tau_\perp$). Throughout the frequency range, we observed a similar behavior of the symmetric and antisymmetric components of the ST-FMR spectra as shown in figure 2(a). The amplitude of the spectra decreases with increasing higher frequency due to decrease in the precession cone angle of magnetization precession. A typical $V_{mix}$ of STO/2DEG/NiFe sample measured at 5 GHz at $\phi = 0°$, as shown figure 2(b) clearly showing the difference in the amplitude of the symmetric and antisymmetric spectra at opposite magnetic field. The effective magnetization, $4\pi M_{eff}$, of STO/2DEG/NiFe sample is estimated from fitting $f$ vs $H_{res}$ plot with in-plane Kittel equation[22].

$$f = \frac{\gamma}{2\pi}\sqrt{(H_{res} + H_k)(H_{res} + H_k + 4\pi M_{eff})} \tag{3}$$

Where, $H_k$ and $\gamma$ are the effective in plane magnetic anisotropic field and gyromagnetic ratio, respectively. The estimated value of $4\pi M_{eff}$ and $\frac{\gamma}{2\pi}$ are $951\ mT$ and $0.0288\ GHz/mT$, respectively. The broadening of ST-FMR has linear relation with frequency as shown figure 2 (d). The Gilbert damping parameter $\alpha$ which depends on linewidth $\Delta H$, is estimated using,

$$\Delta H = \Delta H_o + \frac{2\pi f}{\gamma}\alpha, \tag{2}$$

Also $\Delta H_o$ and $\alpha$ is estimated using linear fitting as shown in equation (2), is the inhomogeneous linewidth broadening and Gilbert damping factor which is independent of $f$, and depends on the sample quality. The value of Gilbert damping factor obtained is 0.007.

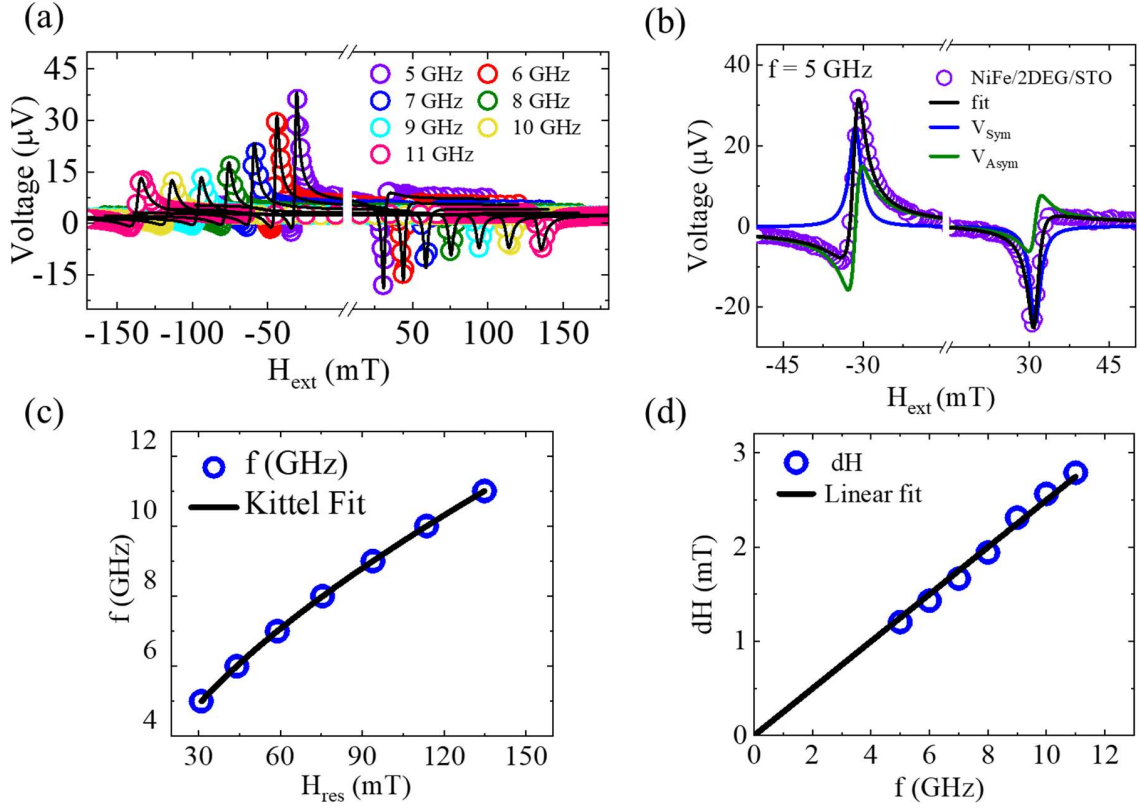

Figure 2. Frequency-dependent ST-FMR measurements: (a) Lorentzian fitting of $V_{mix}$ voltage for 5 to 11 GHz frequency (b) Lorentzian fitting of $V_{mix}$ voltage for both negative and positive applied magnetic field at 5 GHz where $V_{Sym}$ and $V_{Asym}$ represents the symmetric and antisymmetric part of the spectra (c) Linear fitting of fullwidth half maxima with frequency (d) Kittel fit of frequencies ranging from 5 to 11 GHz at $\phi = 0°$ and the corresponding $M_{eff}$.

In a two-dimensional electron gas (2DEG) interface exhibiting substantial spin-orbit coupling (SOC), the flow of a charge current ($J_c$) gives rise to a transverse spin current $J_S$. This $J_S$ exerts a torque on the magnetization vector of the adjacent layer NiFe. The exerted torque can perform ferromagnetic resonance in the adjacent layer while out balancing damping like torque ($\tau_{DL}$) in the external applied magnetic field $H_{ext}$. The magnetic material NiFe exhibits a change in resistance when $J_c$ flows through it, known as anisotropic magnetoresistance (AMR). This effect arises from spin-orbit coupling and depends on the angle between the $H_{ext}$ and the applied $J_c$. In-plane angular-dependent ST-FMR measurements were performed to fully comprehend the symmetry torques concerning the applied external magnetic field ($H_{ext}$). It has been shown that when both the polarization $\hat{p}$ of spin current and the rf field $H_{rf}$ are parallel to y axis, and are transverse to nonmetal/ferromagnetic material, the amplitudes of both the symmetric and antisymmetric components are known to be proportional to sin(2ϕ)cos(ϕ)[23]. This spin polarization configuration is called conventional spin polarization spin current and here $\sin(2\phi)$ contribution is due to AMR of the magnetic layers, while the $cos(\phi)$ contribution describe the angular dependence of the torque amplitude exerted on magnetization by $I_{rf}(\propto J_c)$. So, the overall angular dependence of symmetric and antisymmetric are proportional to $\sin(2\phi)\cos(\phi)$. For Pt/Py sample both symmetric and antisymmetric component of $V_{mix}$ are proportional to $\sin(2\phi)\cos(\phi)$ as shown in figure S5, which can be attributed to very low magnetization of platinum. If the symmetric and antisymmetric component of $V_{mix}$ has additional dependence including the $\sin(2\phi)\cos(\phi)$. This shows there are additional torque components working on the FM magnetization vector. These additional torques arises due when $\hat{p}$ and $H_{rf}$ contain $x$ and $z$ components and this spin polarization configuration is called unconventional spin polarization. The angular dependence of symmetric and antisymmetric components of $V_{mix}$ is obtained at $5\ GHz$ for 25 minutes argon milled STO/2DEG/NiFe sample as shown in figure 2 (a) and (b) respectively. The symmetric component of is fitted using the following equation[24,25]

$$V_{Sym} = S_x sin(2\phi)Sin(\phi) + S_y sin(2\phi)cos(\phi) + S_z sin(2\phi) \qquad (4)$$

Where the $S_x$, $S_y$ and $S_z$ are the weightage of $sin(2\phi)sin(\phi)$, $sin(2\phi)cos(\phi)$ and $sin(2\phi)$. Similarly, the antisymmetric component is fitted using the following equation.

$$V_{Asym} = A_x sin(2\phi)sin(\phi) + A_y sin(2\phi)cos(\phi) + A_z sin(2\phi) \qquad (5)$$

Where the $A_x$, $A_y$ and $A_z$ are the weightage of $sin(2\phi)sin(\phi)$, $sin(2\phi)cos(\phi)$ and $sin(2\phi)$. The angular dependence has additional components in addition to $sin(2\phi)cos(\phi)$ unlike in case of Pt/NiFe[26]. This indicates the breaking of twofold $(180° + \phi)$ and mirror symmetry $(180° - \phi)$. These additional contributions may arise due to non-uniform microwave current flow in the devices, since NiFe has much lower resistivity compared to the 2DEG formed at STO interface, that may give rise to the additional angular dependent components in the ST-FMR spectra.

We note that, in our devices, the angular dependence of symmetric and antisymmetric components is not purely $sin(2\phi)cos(\phi)$. Table in figure 3 (c) shows the weightage of the different symmetric and antisymmetric parts of the spectra at 5 GHz in the case of the STO/2DEG/NiFe sample. Here the symmetric component has 16.79% $sin(2\phi)cos(\phi)$ dependence, 80.71% arising from $sin(2\phi)sin(\phi)$ dependence with the rest 2.50% arising from $sin(2\phi)$. Meanwhile, its antisymmetric component has 2.93%, 60.58%, and 36.49% contributions from $sin(2\phi)cos(\phi)$, $sin(2\phi)sin(\phi)$, and $sin(2\phi)$ dependence, respectively. The large weightage of $sin(2\phi)sin(\phi)$ present in both symmetric and antisymmetric spectra in our sample, unlike Pt/NiFe where this component has nearly zero percentage weightage due to only, SHE presents in Pt, here is shows the presence of strong spin orbit coupling present at interface of STO and NiFe.

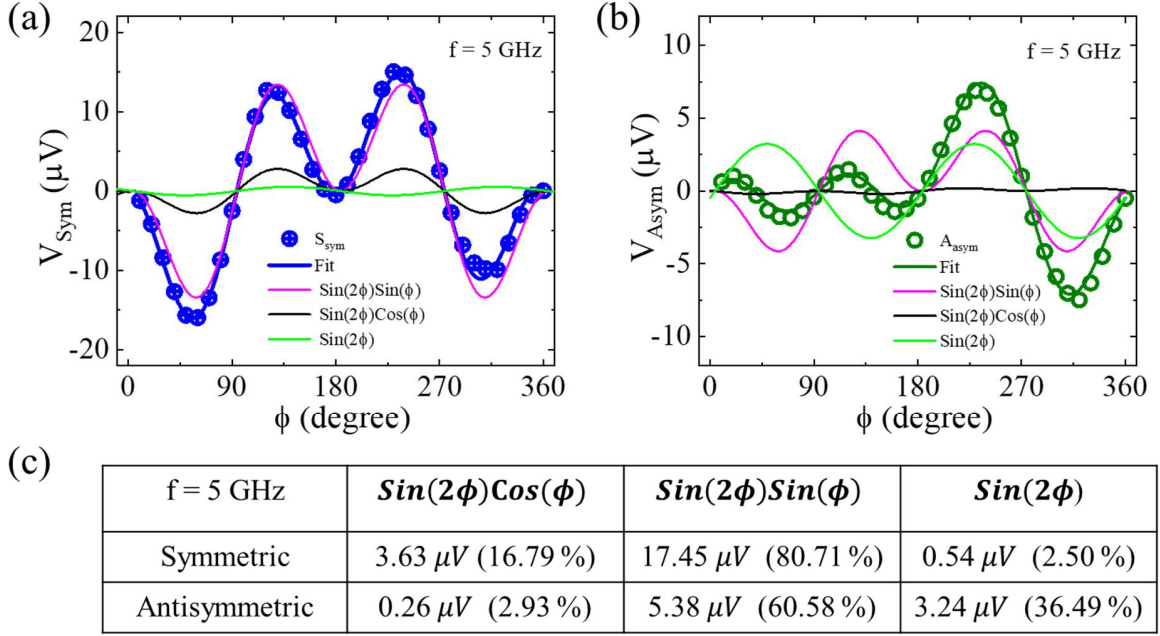

Figure 3. In plane angle-dependent ST-FMR measurements: (a, b) Extracted component from Lorentzian fitting of $V_{mix}$ data for in plane angle ($\phi$) sweep (a) Symmetric contribution $V_{Sym}$ and its fitted components as function of in plane angle between applied $I_{rf}$ and the external magnetic field $H_{ext}$ (b) Antisymmetric contribution $V_{Asym}$ and its component with the in plane angle ($\phi$) (c) Table showing the percentage contribution of different type of torques in symmetric and in antisymmetric part of $V_{mix}$ voltage.

To better understand the spin polarization dependence of symmetric and antisymmetric component of the ST-FMR signal, we have performed a simulation using MuMax[3] [27]. The simulation uses the Landau-Lifshitz-Gilbert-Slonczewski equation [28,18]

$$\frac{d\vec{m}}{dt} = -\gamma \vec{m} \times \vec{B}_{eff} + \alpha \vec{m} \times \frac{d\vec{m}}{dt} + \frac{\gamma |J_C| \hbar \theta_{SH}}{2e t_{FM} \mu_0 M_S}[\vec{m} \times (\vec{\sigma} \times \vec{m}) + \xi_{FL}\, \vec{m} \times \vec{\sigma}] \qquad (6)$$

where $\vec{m}$ is the normalized magnetization, $\vec{B}_{eff}$ is the effective magnetic field and $\alpha$ is Gilbert damping factor, $t_{FM}$ is the thickness of FM layer, $\mu_0 M_S$ is the saturation magnetization. Here $\tau_{ST}[\vec{m} \times (\vec{\sigma} \times \vec{m})]$ and $(\tau_{ST} \times \xi)\vec{\sigma} \times \vec{m}$ are damping-like and field-like torques respectively. We simulate NiFe/Pt and NiFe/STO ferromagnet (FM) / nonmagnet (NM) bilayers.

We use $NiFe$ with a width of 160 nm and thickness of 5 nm. Simulation parameters are shown in the table in figure 4 (b). The applied current density $J_C$ is in x direction exerting a magnetic field $B_{oe}$ on the NiFe and due to spin Hall effect in NM, generated spin current will exert SOT in NiFe. The charge to spin conversion efficiency (SHA) is taken as 100% for NiFe/STO and and 5% in case of NiFe/Pt in the simulation. We have performed ST-FMR simulation for NiFe/Pt sample using conventional case of spin polarization, with $\hat{p}$ of spin current and $H_{rf}$, arising from applying the $I_{rf}$ in $x$ direction, both are in $y$ direction, transverse to the nonmetal/ferromagnetic material. The simulation results are shown in (Supplementary figure S5) the angular dependence of ST-FMR spectra for both symmetric and antisymmetric spectra as $sin(2\phi)cos(\phi)$.

As for STO($Ar^+$)/NiFe sample the in-plane angle dependence of symmetric and antisymmetric components was not proportional to the $sin(2\phi)cos(\phi)$ suggesting presence of additional torque. We have simulated this using unconventional spin polarization case by keeping the $H_{rf}$ field in $y$ direction and taking polarization as $(1, -0.3, -0.05)$. The presence of x and z components in spin polarization invites additional torques due to which the symmetric and antisymmetric component was no longer showing the $sin(2\phi)cos(\phi)$ dependence. The studies have shown that angular dependence other than $sin(2\phi)sin(\phi)$ appear in presence of anisotropic spin relaxation due to Rashba-type spin orbit field or due to non-uniform distribution of $I_{rf}$. The simulation shows symmetric part has additional terms $sin(2\phi)sin(\phi)$ and $sin(2\phi)$, arising due to damping like torque in x direction and field like torque in z direction due to Oersted field. Similarly, there are additional terms $sin(2\phi)sin(\phi)$ and $sin(2\phi)$, which arises from field like torque in x direction and damping like torque in z direction.

The simulation shows large dependence of symmetric and antisymmetric on $sin(2\phi)sin(\phi)$ in unconventional spin polarization case, unlike conventional spin polarization case of Pt/Py where this is completely zero. Since in simulation $I_{rf}$ distribution is same for both, we can attribute this effect to anisotropic spin diffusion in ferromagnet due to strong Rashba like SOC present at the substrate and NiFe material interface.

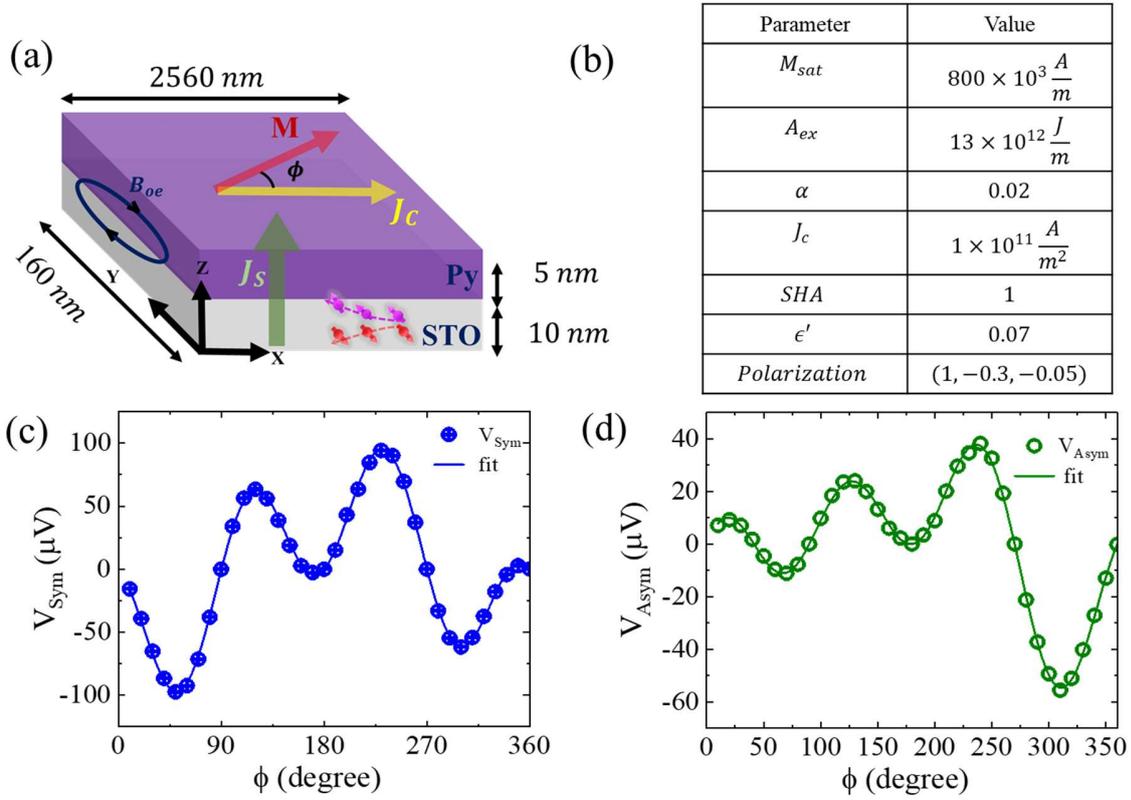

*Figure 4. In plane angle-dependent ST-FMR data obtained from MuMax³: In plane angle dependent $V_{mix}$ from ST-FMR simulation using MuMax³ at 5 GHz (a) schematic showing the geometry of the sample, direction of charge current, spin current and applied external magnetic field (b) table showing the parameters used in ST-FMR simulation (c) symmetric component within plane angle (d) antisymmetric component within plane angle.*

In summary, we have demonstrated unconventional spin polarization at the argon milled STO

interface interfacing with NiFe. The symmetric and antisymmetric components of rectification voltage revealing the breaking of spin orbit torque symmetries. The micromagnetic simulation study under the similar experimental conditions further confirms the presence of the unconventional spin polarization at the interface. Our results emphasize the potential of modifying material surface properties through ion milling to achieve efficient charge to spin conversion, with promising of development of a new kind of oxide-based spintronics.

**Data availability statement**

The data that support the findings of this study are available from the corresponding author upon reasonable request.

**References:**

(1) Rojas-Sánchez, J.-C.; Oyarzún, S.; Fu, Y.; Marty, A.; Vergnaud, C.; Gambarelli, S.; Vila, L.; Jamet, M.; Ohtsubo, Y.; Taleb-Ibrahimi, A. Spin to Charge Conversion at Room Temperature by Spin Pumping into a New Type of Topological Insulator: α-Sn Films. *Phys. Rev. Lett.* **2016**, *116* (9), 96602.

(2) Everhardt, A. S.; Mahendra, D. C.; Huang, X.; Sayed, S.; Gosavi, T. A.; Tang, Y.; Lin, C.-C.; Manipatruni, S.; Young, I. A.; Datta, S. Tunable Charge to Spin Conversion in Strontium Iridate Thin Films. *Phys. Rev. Mater.* **2019**, *3* (5), 51201.

(3) Shashank, U.; Deka, A.; Ye, C.; Gupta, S.; Medwal, R.; Rawat, R. S.; Asada, H.; Renshaw Wang, X.; Fukuma, Y. Room-Temperature Charge-to-Spin Conversion from Quasi-2D Electron Gas at SrTiO3-Based Interfaces. *Phys. status solidi (RRL)–Rapid Res. Lett.* **2022**, 2200377.

(4) Sánchez, J. C. R.; Vila, L.; Desfonds, G.; Gambarelli, S.; Attané, J. P.; De Teresa, J. M.; Magén, C.; Fert, A. Spin-to-Charge Conversion Using Rashba Coupling at the Interface between Non-Magnetic Materials. *Nat. Commun.* **2013**, *4* (1), 2944.

(5) Hellman, F.; Hoffmann, A.; Tserkovnyak, Y.; Beach, G. S. D.; Fullerton, E. E.; Leighton, C.; MacDonald, A. H.; Ralph, D. C.; Arena, D. A.; Dürr, H. A. Interface-Induced Phenomena in Magnetism. *Rev. Mod. Phys.* **2017**, *89* (2), 25006.

(6) Chen, J.; Wu, K.; Hu, W.; Yang, J. Spin–Orbit Coupling in 2D Semiconductors: A Theoretical Perspective. *J. Phys. Chem. Lett.* **2021**, *12* (51), 12256–12268.

(7) Manchon, A.; Koo, H. C.; Nitta, J.; Frolov, S. M.; Duine, R. A. New Perspectives for Rashba Spin–Orbit Coupling. *Nat. Mater.* **2015**, *14* (9), 871–882.

(8) Zhang, J.; Zhang, J.; Chi, X.; Hao, R.; Chen, W.; Yang, H.; Zhu, D.; Zhang, Q.; Zhao, W.; Zhang, H. Giant Efficiency for Charge-to-Spin Conversion via the Electron Gas at the LaTiO 3+ δ/SrTiO 3 Interface. *Phys. Rev. B* **2022**, *105* (19), 195110.

(9) Yang, H.; Zhang, B.; Zhang, X.; Yan, X.; Cai, W.; Zhao, Y.; Sun, J.; Wang, K. L.; Zhu, D.; Zhao, W. Giant Charge-to-Spin Conversion Efficiency in Sr Ti O 3-Based Electron Gas Interface. *Phys. Rev. Appl.* **2019**, *12* (3), 34004.

(10) Wang, Y.; Ramaswamy, R.; Motapothula, M.; Narayanapillai, K.; Zhu, D.; Yu, J.; Venkatesan, T.; Yang, H. Room-Temperature Giant Charge-to-Spin Conversion at the SrTiO3–LaAlO3 Oxide Interface. *Nano Lett.* **2017**, *17* (12), 7659–7664.

(11) Takeuchi, Y.; Hobara, R.; Akiyama, R.; Takayama, A.; Ichinokura, S.; Yukawa, R.; Matsuda, I.; Hasegawa, S. Two-Dimensional Conducting Layer on the SrTi O 3 Surface Induced by Hydrogenation. *Phys. Rev. B* **2020**, *101* (8), 85422.

(12) Huang, C.; Milletari, M.; Cazalilla, M. A. Spin-Charge Conversion in Disordered Two-Dimensional Electron Gases Lacking Inversion Symmetry. *Phys. Rev. B* **2017**, *96* (20), 205305.

(13) Silvestrov, P. G.; Zyuzin, V. A.; Mishchenko, E. G. Mesoscopic Spin-Hall Effect in 2D Electron Systems with Smooth Boundaries. *Phys. Rev. Lett.* **2009**, *102* (19), 196802.

(14) Gorini, C.; Sheikhabadi, A. M.; Shen, K.; Tokatly, I. V; Vignale, G.; Raimondi, R. Theory of Current-Induced Spin Polarization in an Electron Gas. *Phys. Rev. B* **2017**, *95* (20), 205424.

(15) Browning, N. D.; Buban, J. P.; Moltaji, H. O.; Pennycook, S. J.; Duscher, G.; Johnson, K. D.; Rodrigues, R. P.; Dravid, V. P. The Influence of Atomic Structure on the Formation of Electrical Barriers at Grain Boundaries in SrTiO 3. *Appl. Phys. Lett.* **1999**, *74* (18), 2638–2640.

(16) Iglesias, L.; Sarantopoulos, A.; Magén, C.; Rivadulla, F. Oxygen Vacancies in Strained SrTiO 3 Thin Films: Formation Enthalpy and Manipulation. *Phys. Rev. B* **2017**, *95* (16), 165138.

(17) Li, Y.; Phattalung, S. N.; Limpijumnong, S.; Kim, J.; Yu, J. Formation of Oxygen Vacancies and Charge Carriers Induced in the N-Type Interface of a LaAlO 3 Overlayer on SrTiO 3 (001). *Phys. Rev. B* **2011**, *84* (24), 245307.

(18) Liu, L.; Moriyama, T.; Ralph, D. C.; Buhrman, R. A. Spin-Torque Ferromagnetic Resonance Induced by the Spin Hall Effect. *Phys. Rev. Lett.* **2011**, *106* (3), 36601.

(19) Liu, L.; Pai, C.-F.; Li, Y.; Tseng, H. W.; Ralph, D. C.; Buhrman, R. A. Spin-Torque Switching with the Giant Spin Hall Effect of Tantalum. *Science (80-. ).* **2012**, *336* (6081), 555–558.

(20) MacNeill, D.; Stiehl, G. M.; Guimaraes, M. H. D.; Buhrman, R. A.; Park, J.; Ralph, D. C. Control of Spin–Orbit Torques through Crystal Symmetry in WTe2/Ferromagnet Bilayers. *Nat. Phys.* **2017**, *13* (3), 300–305.

(21) Stiehl, G. M.; MacNeill, D.; Sivadas, N.; El Baggari, I.; Guimarães, M. H. D.; Reynolds, N. D.; Kourkoutis, L. F.; Fennie, C. J.; Buhrman, R. A.; Ralph, D. C. Current-Induced Torques with Dresselhaus Symmetry Due to Resistance Anisotropy in 2D Materials. *ACS Nano* **2019**, *13* (2), 2599–2605. https://doi.org/10.1021/acsnano.8b09663.

(22) Kittel, C.; McEuen, P. Introduction to Solid State Physics, Vol 8 Wiley New York. **1976**.